\documentclass[aps,prl,10pt,twocolumn,floatfix,superscriptaddress]{revtex4-2}
\usepackage{amsfonts,amsmath,graphicx,braket,bm}
\usepackage[utf8]{inputenc}

\begin{document}
\title{Multiscale Quantum Approximate Optimization Algorithm}
\author{Ping Zou}
\email{zouping@m.scnu.edu.cn}
\affiliation{Guangdong Provincial Key Laboratory of Quantum Engineering and Quantum Materials, School of Information and Optoelectronic Science and Engineering, South China Normal University, Guangzhou 510006, China}
\date{\today} 

\begin{abstract}
The quantum  approximate optimization algorithm (QAOA) is one of the canonical algorithms designed to find approximate solutions to combinatorial optimization problems in  current noisy intermediate-scale quantum (NISQ) devices.
It is an active  area of research  to exhibit its speedup over classical algorithms. The performance of the QAOA at low depths is limited, while the QAOA at higher depths is constrained by the current techniques. We propose a new version of QAOA that incorporates the capabilities of QAOA and the real-space renormalization group transformation, resulting in enhanced performance.   Numerical simulations demonstrate that our algorithm can provide accurate solutions for certain randomly generated instances utilizing QAOA at low depths, even at the lowest depth. The algorithm is suitable for NISQ devices to exhibit a quantum advantage.
\end{abstract}

\maketitle

\paragraph*{Introduction.} 
Variational quantum algorithms (VQAs) have gained significant research attention due to their potential to provide quantum advantages on near-term noisy intermediate-scale quantum (NISQ) computers \cite{Bharti2022, peruzzo2014, debnath2016, barends2016}.  The
quantum approximate optimization algorithm (QAOA) is  one of the first  proposals of VQAs designed to solve combinatorial optimization problems \cite{farhi2014}.
It has been widely studied over the years, both theoretically and empirically \cite{jiang2017, morales2018variational, Akshay2020, lloyd2018, juan2023, L.Zhou 2020, Bravyi2020}. QAOA optimizes an ansatz quantum  state prepared  through alternating evolutions between a simple mixing Hamiltonian  and the problem Hamiltonian.  The depth of QAOA is a crucial parameter that influences the measurement outcomes of local operators. If  the depth is small,  a local qubit does not ``see'' the entire graph, leading to limited performance \cite{Bharti2022, farhi2020a, farhi2020b}. 
It has also  shown that a fixed depth QAOA exhibits reachability
deficits \cite{Akshay2020, Bharti2022}. Although many methods have been proposed to improve the performance of QAOA, increasing the depth is a widely accepted and effective strategy \cite{L.Zhou 2020}. However, the  depth is restricted by two factors. One is the
limited coherence time of the current NISQ devices, 
which the preparation of ansatz states  and measurements must be implemented within. 
The other is that classical optimization with
too many parameters may suffer from \textit{barren plateaus} \cite{mcclean2018barren}.

To  overcome these shortcomings, we  propose a new version of the algorithm: multiscale QAOA (MQAOA). It is powered by a real-space renormalization group (RG) transformation. Based on the truncation of Hilbert space, a real-space RG method  is first proposed for the numerical  computation of strongly interacting  quantum systems 
defined on a lattice\ \cite{wilson1975}, then it is improved
by the density matrix renormalization group (DMRG) algorithm
 \cite{white1992, white1993}.  The DMRG algorithm has been proven to be a highly reliable, precise and versatile numerical method for the  calculation of statics, dynamics or thermodynamic quantities  in quantum systems defined on low dimensional lattices \cite{sch2005}.
Later, to investigate quantum systems on a $D$-dimensional lattice or a scale invariant system where DMRG fails, a further extending method named  entanglement renormalization  is proposed \cite{G.vidal 2007, G.vidal 2008, G.vidal 2009}. 
We propose a  real-space RG  transformation for the quantum system defined on a graph, 
and merge it with  QAOA to form the MQAOA. By utilizing a feedback loop of QAOA and the real-space RG transformation, MQAOA  significantly improves the performance  compared to the standard QAOA. Numerical testing  demonstrates that MQAOA can provide accurate solutions  for  specific randomly generated problems  by exploiting QAOA of low depths.

\paragraph*{Quantum approximate optimization algorithm.}
Considering binary combinatorial optimization problems such as MaxCut\cite{farhi2014}, Max-2-SAT\cite{Akshay2020},  and QUBO\cite{Koch2014}, they are all NP-hard and often used as 
benchmarks for  QAOA. These problems can be described by a quantum system defined on a  graph $G=(V, E)$ composed by a set of   $N$ vertices $V$ labeled by integers $j=1, 2, \cdots, N$ and a set of edges $E$. 
The bit in the binary string relevant to a problem is represented by 
qubit on the corresponding vertex. The cost function of a problem is encoded into a Hamiltonian written as:
\begin{equation}
H_z = \sum_{i<j}J_{ij} \sigma_i^z  \sigma_j^z +\sum_i h_i\sigma_i^z,  \label{ising}
\end{equation}
where $J_{ij}$ is the  coupling coefficient between qubit $i$ and $j$, $h_i$ is the external field,  and $\sigma_i^z $ is the Pauli $Z$ operator applied to  qubit $i$. 
In the standard formula, 
another mixing Hamiltonian   $H_x = \sum_i \sigma_i^x $ is utilized. 
For a  depth-$p$ QAOA, a quantum computer is used to prepare  the variational state as
\begin{equation}
\ket{\psi_p(\boldsymbol{\beta}, \boldsymbol{\gamma})} =
\prod_{k=1}^{p}e^{-iH_x\beta_k/\hbar}e^{-iH_z\gamma_k/\hbar}\ket{\psi_0},
\end{equation}
where evolution periods $\beta_i$ and $\gamma_i$ ($i=1, 2, \cdots, p$) are variational parameters,  and the initial state $\ket{\psi_0}=\ket{+}^{\otimes N}$ is the tensor product of $N$ single qubit states  $\ket{+}=(\ket{0}+\ket{1})/\sqrt{2}$, with $\ket{0}, \ket{1}$ 
the eigenstates of $\sigma_z$ corresponding to eigenvalues $\pm 1$.
After  the variational state is prepared, local measurements are performed 
to obtain the expected energy of $H_z$, 
$F(\boldsymbol{\beta},\boldsymbol{\gamma})=
\bra{\psi_p(\boldsymbol{\beta}, \boldsymbol{\gamma})}
H_z\ket{\psi_p(\boldsymbol{\beta}, \boldsymbol{\gamma})}$.
$F(\boldsymbol{\beta},\boldsymbol{\gamma})$ is the objective cost function of a 
classical optimization method to search its minimum with optimal parameters $\boldsymbol{\beta}^*$
and $\boldsymbol{\gamma}^*$.  The  search can be executed by  a simplex
or gradient-based optimization with initial guessing  parameters.
The  standard QAOA  presented above is inspired by the Trotter approximation  of  the quantum adiabatic algorithm \cite{farhi2014}. However,  it can learn  to utilize nonadiabatic mechanisms to overcome the difficulties associated with vanishing spectral gaps  through optimization \cite{L.Zhou 2020}. There is a feature of geometric locality  for the QAOA ansatz state \cite{farhi2014, Bravyi2020}. For a depth-$p$ QAOA, the expected value associated with edge $\langle j,k\rangle$ only involves qubits $j$,  $k$  and those vertices whose distances on the graph from $j$ or $k$  are less than or equal to $p$.   If it is desired to establish connections between qubits with long distances using only  QAOA at low depths, it may be advantageous to employ a real-space RG transformation.

\paragraph*{Real-space RG on graphs.}
 For  a quantum system defined on a lattice $\mathcal{L}$,  blocks of its neighboring sites are coarse-grained into single sites of a new effective lattice $\mathcal{L}^\prime$.
  The Hilbert space of the block $\mathbb{V}_{\mathcal{B}}$
is truncated into a space $\mathbb{V}_s$ for site $s\in \mathcal{L}^\prime$,  which is a key step
for the real-space RG transformation.
The dimensional truncation of the coarse-graining is characterized by an  isometry $\omega$ \cite{G.vidal 2007},  
\begin{equation}
	\omega : \mathbb{V}_s \rightarrow \mathbb{V}_{\mathcal{B}}, \quad \quad   \omega ^\dagger \omega=I_{\mathbb{V}_s}, 
	\quad \quad  \omega\omega^\dagger=P,
\end{equation}
where $I_{\mathbb{V}_s}$ is the identity operator in $\mathbb{V}_s$, and $P$ is a projector onto the subspace of $\mathbb{V}_{\mathcal{B}}$ that is preserved by the coarse-graining.
The effective  Hamiltonian $H^\prime$ after the RG transformation is given by
\begin{equation}
	H^\prime = W^\dagger H W,
\end{equation}
where $H$ is the Hamiltonian defined on lattice $\mathcal{L}$, $W=\otimes \omega^{(k)}$ is the tensor product of all isometries,  and $\omega^{(k)}$ is the isometry for  block $k$. There are two noteworthy
coarse-graining schemes, the DMRG and entanglement renormalization.
DMRG algorithm uses a general decimation prescription. 
For a wave function $\ket{\psi}$, typically a ground state, the process involves calculating the reduced density matrix $\rho_b$ for  each block,  determining its eigensystem through exact diagonalization and keeping $m$ orthonormal eigenstates with the largest associated eigenvalues as the reduced basis.
Thus, the matrix representation of an isometry $\omega$   consists of 
$m$ eigenstates as columns. Here $m$ is the dimension of the 
new site, and it determines the truncation error \cite{white1992, G.vidal 2007, sch2005, sch2011}.
Such a choice minimizes the  Frobenius norm of the difference between $\rho_b$ and its projection onto an $m$-dimensional subspace $\left \Vert \rho_b-P \rho_b P\right\Vert$ \cite{white1992, sch2011}.
While in the entanglement renormalization  scheme proposed by G. Vidal \cite{G.vidal 2007}, before  the coarse-graining step performed as in  DMRG algorithm,   short-range entanglement localized near the boundary of a block is eliminated by  using disentanglers, then the original state $\rho_b$ is replaced by a partially disentangled state $\tilde{\rho}_b$.

Considering the character of the problem Hamiltonian of QAOA,  we propose a  new coarse-graining transformation for a quantum system defined on a graph $G$. We first partition neighboring vertices  into blocks by finding a maximal matching $M$ of graph  $G$.   A matching is a subset of edges in which no vertex occurs more than once. A maximal matching  is a matching that cannot be enlarged by adding an edge, and  it is usually not unique. We export a greedy algorithm implementation for the searching from the Python package NetworkX \cite{Hagberg2008}.
Each pair of  vertices connected by an edge in $M$  is  grouped into a block. By coarse-graining, 
each block is transformed into an effective vertex with one qubit defined on it. 
The graph is then transformed into an effective one with  much fewer vertices. 
Unlike the  numerical simulation methods, our  scheme  is a hybrid quantum  classical algorithm. 
The reduced density matrix $\rho_b$  for each block is reconstructed through measurements of the quantum computer after the ansatz state with optimal parameters prepared. 
Concretely,  the density matrix $\rho^{(k)}$  of block $k$ consisting of two adjacent sites, $i$ and $j$,  can be reconstructed by measuring the expected values of 15 local operators $\langle\sigma_i^{\alpha}\sigma_j^{\beta}\rangle$: $\rho^{(k)} = \sum_{\alpha, \beta}\langle\sigma_i^{\alpha}\sigma_j^{\beta}\rangle  \sigma_i^{\alpha}\sigma_j^{\beta}$, 
where $\alpha, \beta= 0,1,2,3$, $\sigma_i^0$ denotes
the identity operator on qubit $i$, and the others are the corresponding Pauli operators.

After reconstructing  the density matrix for block $k$, 
a $2$-dimensional subspace is built with basis $\{\ket{v_0}, \ket{v_1}\}$.  
The corresponding isometry is  a matrix with  the two basis vectors as columns, and the projector  is $P=\ket{v_0}\bra{v_0}+\ket{v_1}\bra{v_1}$. 
The basis is selected to minimize $\left\Vert\rho^{(k)}-P \rho^{(k)} P\right \Vert$,  similar 
as in DMRG.  However, there is an additional condition that requires three operators $\sigma_i^z$, $\sigma_j^z$ and $\sigma_i^z\sigma_j^z$ to be diagonal with respect to the new basis vectors.   The constraint is included to guarantee that the resulting effective Hamiltonian retains an Ising form, as presented in Eq. (\ref{ising}). Afterward, QAOA can be employed to solve the effective Hamiltonian once more. Considering the factor that the identity operator is naturally diagonal, then equivalently, four operators $(I\pm \sigma_i^z)(I\pm \sigma_j^z)/4$ should be diagonal with respect to the basis. A simple analysis reveals that there are  two possible scenarios. In the first scenario, two of the four components in the column vector $\ket{v_0}$ are zeros, while the remaining components in $\ket{v_1}$ are zeros. For example, the first two components of $\ket{v_0}$ and the last two  of $\ket{v_1}$ are zeros, which implies that qubit $i$ is in either  definite state of $\ket{0}$ or $\ket{1}$,  while qubit $j$ is in a superposition state. The second scenario is that one component of $\ket{v_0}$ is zero while the other three components of $\ket{v_1}$  are zeros.  We take the first scenario  and refer to \cite{supp} for a detailed calculation.

\begin{figure}[tbp]
	\centering
	\includegraphics[width=0.5\textwidth]{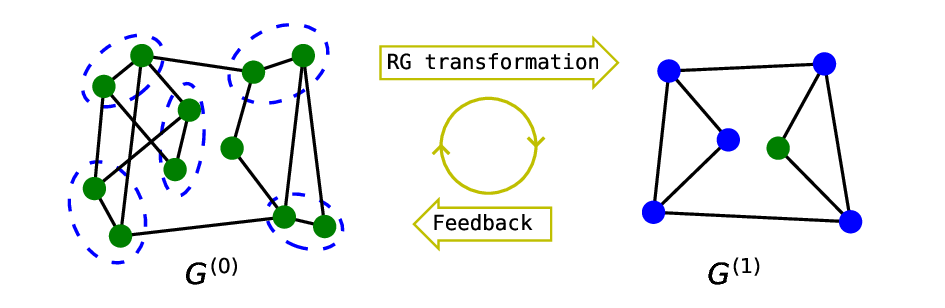}
	\caption{(color online). The workflow of a MQAOA. For a quantum system defined on a graph $G^{(0)}$, pairs of vertices connected by an edge in a maximal matching are grouped.
	After preparing  the QAOA's ansatz state with optimal parameters, a real-space RG transformation is performed to get an effective graph $G^{(1)}$ and the corresponding Hamiltonian $H^{(1)}_z$. A ground state of $H^{(1)}_z$ is then found and used to set the initial state for a new round of QAOA on $G^{(0)}$. The loop continues until the expected energy of $H^{(0)}_z$ converges. The ground state of $H^{(1)}_z$ can also be obtained  through the same procedure.
	} \label{TMEV0}
\end{figure}

\paragraph*{Multiscale QAOA.}
For a graph $G^{(0)}$ with the corresponding Hamiltonian $H_z^{(0)}$,  the MQAOA consists of the following steps, which 
is visualized in Fig. {\ref{TMEV0}.
\begin{enumerate}
	\item Find a maximal matching $M$ of $G^{(0)}$.
	\item  Perform the QAOA of low depth on  $G^{(0)}$ to find optimal parameters.  In the first round, the initial  state is  $\ket{\psi_0}=\ket{+}^{\otimes N}$.
	\item  Prepare the variational state with  optimal parameters,  and implement measurements to reconstruct reduced density matrix for each pair of qubits connected by edges in $M$, then perform the RG transformation  to obtain a new coarse-grained graph $G^{(1)}$ and the corresponding Hamiltonian $H_z^{(1)}$.
	\item Find a  ground state $\ket{\psi^{(1)}}$ of $H_z^{(1)}$. 
	\item  Prepare the quantum state $W\ket{\psi^{(1)}}$  on  $G^{(0)}$, which is the initial guessing state  for a new round of QAOA.
	\item Repeat steps ($2$)-($5$) until  the expected energy  of $H_z^{(0)}$ converges.
\end{enumerate}
In step ($4$), to find  the ground state of $H_z^{(1)}$ on  graph $G^{(1)}$, 
a new MQAOA  can be carried out  to get an effective graph $G^{(2)}$  and the corresponding Hamiltonian $H_z^{(2)}$  through  another coarse-grained step. 
By iterating, a sequence of Hamiltonians $\{ H_z^{(n)} \}$ is generated until a particular value of $n_0$ is reached, for which $H_z^{(n_0)}$ can be effectively solved by either a quantum or classical computer. If there is more than one  ground state for $H_z^{(n_0)}$ due to its symmetry,  a product state is selected. The selection of a symmetry broken state   ensures that  the algorithm overcomes the limitation from symmetry protection as proposed in \cite{Bravyi2020}. 
The initial state  $W\ket{\psi^{(1)}}$ in each round is still a tensor product of local states.  Depending on  the ground state of the next level effective Hamiltonian in the coarse-graining sequence, the qubits  connected by an edge in $M$  are prepared in one of the two basis vectors, both of which are  product states, and the qubits not connected by edges in $M$ are prepared in state  $\ket{0}$ or $\ket{1}$. For only single qubit gates are needed, the initial states can be prepared with high precision in NISQ devices.

In the algorithm, there is a competition between the short-distance connection in step ($2$) and  the long-distance connection in step ($4$), which enables MQAOA to break the locality limitation in a standard  QAOA at low depths. Because we need to implement QAOA with  a sequence of iterative Hamiltonians, which have different connection scales within the original graph,  we have termed the algorithm as  multiscale QAOA.

We demonstrate the efficiency of MQAOA using a paradigmatic test:  MaxCut problem.
The corresponding  $N$-qubit problem Hamiltonian is 
\begin{equation}
H_C = \frac{1}{2} \sum_{(i,j)\in E}(I-\sigma_i^z\sigma_j^z).
\end{equation}
It is a specific Ising Hamiltonian represented by Eq. (\ref{ising}). We aim to find the state with maximum energy or the ground 
state of $H_C^\prime = \frac{1}{2} \sum_{(i,j)\in E} \sigma_i^z\sigma_j^z $ equivalently. As usual, the performance of MaxCut problem  is quantified by the approximation 
ratio  $r$ defined as the ratio between the maximum variational energy and the maximum exact energy of  $H_C$. 

We start by investigating the  MaxCut problem on  a cycle
graph,  which is known as the {\sl ring of disagrees}.  It has been 
studied by numerical computation  and theoretical analysis \cite{farhi2014, wang2018quantum, Mbeng1911}.
The optimal approximation ratio achieved by a depth-$p$ QAOA for even $N$ is bounded above by $(2p+1)/(2p+2)$ for all $p<N/2$. The exact solution can be obtained with the depth $p \ge N/2$.  In the run of MQAOA,  each of two adjacent qubits is grouped,  and translational symmetry is  employed, so  the results are applicable for any even integer $N$.    Depth-$1$ and depth-$2$ QAOAs  are both tested, and the effective Hamiltonians are exactly solved  by classical algorithm after one step of RG transformation.   The results  are shown in  Fig. \ref{performance}(a).  We can observe that both QAOA and RG transformation improve the performance in each round, and near exact solution can be achieved in $6$ rounds. 
As anticipated, depth-$2$  QAOA makes convergence faster since applying a  QAOA of higher depth should bring the variational state closer to the ground state. We can  also observe that the performance  ratio of  MQAOA converges to  $1$ much more rapidly than that of the standard QAOA by increasing the depth.
\begin{figure}[tbp]
	\centering
	\includegraphics[scale=0.5]{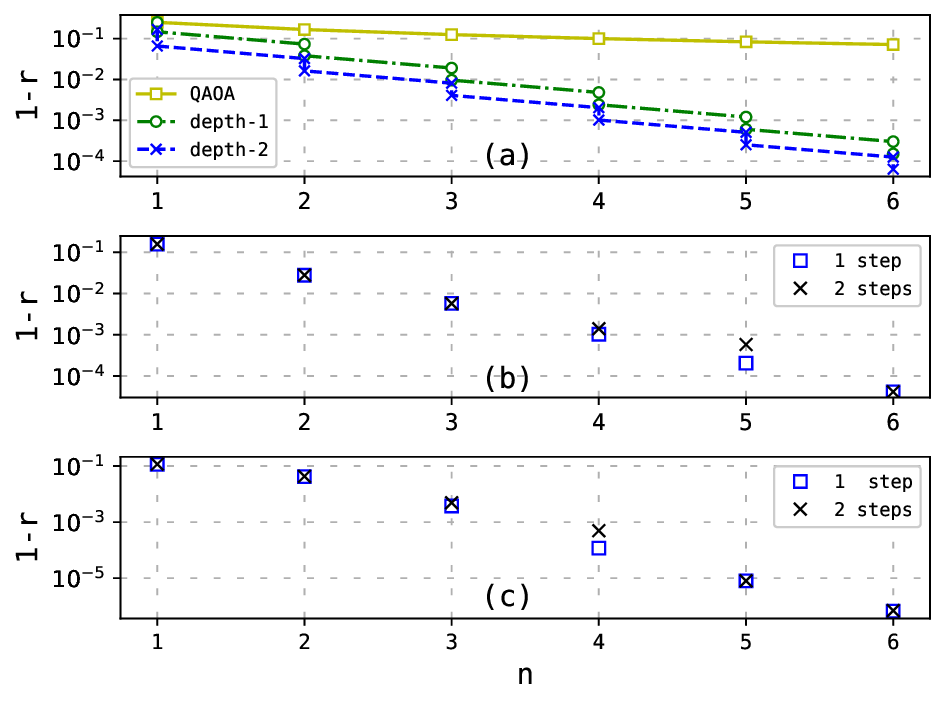}
	\caption{(color online). Performance of the QAOA as measured by the fractional error $1-r$ is plotted as a function of $n$ on a log-linear scale,  where  $r$ is the approximation ratio, $n$ is the number of rounds. (a) For the cycle graph, in each round, both the errors obtained after QAOA (upper points) and  real-space RG transformation (lower points) are shown. In each RG step, both depth-$1$ (green dot dash line) and depth-$2$ (blue dash line) QAOAs are  tested. For comparison, we also plot the downward trend of $1-r$ along with the depth $n$ in the standard QAOA (yellow line). (b) For random $3$-regular graphs with $40$ vertices and (c)  connected \textit{Erdös-Rényi} graphs with $20$ vertices and edge density $\rho=0.1$,  results with one and two steps  of coarse-graining  are  both demonstrated. All data are the mean  of $1-r$  over $100$ randomly generated instances. The exact maximum energies are computed using mixed-integer linear programming \cite{Es2007}.
	} \label{performance}
\end{figure}

We proceed with considering the problems on two typical families of graphs:  random regular graphs
and \textit{Erdös-Rényi} graphs. These graphs are sampled using the NetworkX package.
The performance ratios of MQAOA on these graphs are shown in Fig. \ref{performance}(b-c).  For all graphs in the randomly sampled ensemble, the exact solution  can be obtained by just adopting depth-1 QAOA in each coarse-graining step.  Both one and two  steps of coarse-graining are tested.  All the performance ratios  converge to  $1$ in approximately $6$  rounds, similar to the cycle graphs. 
   
Numerical experiments indicate that the algorithm is prone to get stuck in a suboptimal solution for some graphs when using a depth-$1$ QAOA. 
Increasing the depth of QAOA is an available way to overcome this issue.  Nevertheless,  we have used an alternative method.  It has been discovered by numerical experiments that the performance ratio may rely on  a particular partition scheme.  Thus, several maximal matchings are tested for the graph in each coarse-graining step to achieve a better solution. To obtain various  matchings, the vertices of a graph are  randomly shuffled  before running the matching searching algorithm.  
For Hamiltonian with non-integer couplings, it is necessary to have additional considerations to prevent suboptimal results.  For example,  edges with tiny weights need be excluded before searching for a maximal matching. Because the pair of qubits connected by these edges have little interaction,  the subsystem consisting of such a pair of qubits is almost in the maximally mixed state after the first round of  QAOA at low depths. 
Including these edges in the matching would provide little helpful information for coarse-graining.

\paragraph*{Computational complexity. }
 \begin{figure}[tbp]
	\centering
	\includegraphics[scale=0.5]{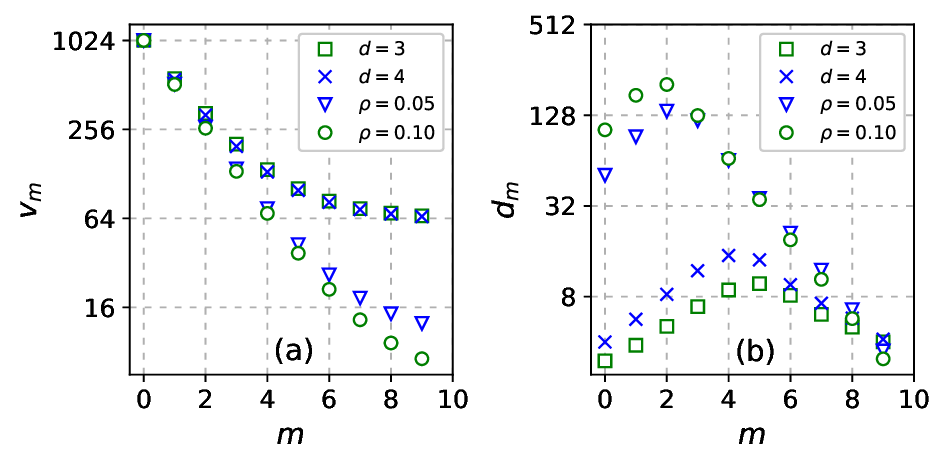}
	\caption{(color online).  (a) The number of vertices $v_m$  and (b) the average degrees $d_m$ of  the effective graphs are plotted along with the number of the iterations of real-space RG transformation on four types of random initial graphs: $d$-regular graphs with degree $d=3$ and $d=4$,  \textit{Erdös-Rényi} graphs with edge density $\rho=0.05$ and $\rho=0.10$.
	The data are mean values over 200 randomly generated graphs.
	} \label{tendency}
\end{figure}
 Through successive  RG transformations, a sequence of  effective  graphs with fewer and fewer nodes is built. Numerical simulation evinces that the number of steps of the RG transformation needed is  $\sim O(\log_2 N)$ to make the graph size small enough to be solved directly, as shown in  Fig. \ref{tendency}(a). We can observe that the decreasing rate of the order of the graph  slows down only in the final few steps, where  the exact solution is  already obtainable.  For \textit{Erdös-Rényi} graphs, especially those with high edge density, the number of vertices is reduced by almost half at each coarse-graining step. For regular graphs, the effective graphs may converge towards a star-shaped graph with a comparatively higher number of vertices, but the ground state of the corresponding Hamiltonian defined on it can be easily solved.  In  Fig. \ref{tendency}(b), we also display the average degrees of effective graphs, which are related to the  vertices   involved in the measurement of the local operators in QAOA at each RG coarse-graining step. In summary, if the number of rounds for each effective Hamiltonian in MQAOA is limited to a constant $K$ to ensure convergence, and the number of different partition schemes is limited to a constant $C$, then the total number of QAOA runs will be $\sim O( N^{\log_2 (CK)})$. Our numerical experiments on graphs with varying types and numbers of vertices indicate that $K=6$ is sufficient. Thus, the MQAOA is an efficient algorithm.

\paragraph*{Conclusion.}
We have proposed the MQAOA to improve the standard QAOA by taking advantage of a real-space RG transformation.  The MQAOA enhances performance significantly.  It can even solve some particular randomly generated instances exactly.  Because just QAOAs at low depths are needed,  it does not suffer from barren plateaus and is suitable for NISQ devices to provide a quantum advantage.

This work was supported by the National Natural Science Foundation of China (Grant No. 62371199) and in part by Guangdong Provincial Key Laboratory(Grant No. 2020B1212060066).

\onecolumngrid

\clearpage
\newpage 

\begin{center}
	\Large\textbf{Supplemental Material for ``Multiscale Quantum Approximate Optimization Algorithm''}
\end{center}

In this supplemental material, we describe the details of  coarse-graining scheme in MQAOA. As noted in the main text, we need to construct a $2$-dimensional subspace from a reduced density matrix $\rho^{(k)}$ of a block consisting of two neighboring qubits, $i$ and $j$, with  a constraint that three operators $\sigma_i^z, \sigma_j^z, \sigma_i^z\sigma_j^z$ should be diagonal with respect to the basis.  There three possible forms for the basis: (a) $\{\ket{0}\ket{\psi_0},\ket{1}\ket{\psi_1} \}$,  (b)
$\{\ket{\phi_0}\ket{0}, \ket{\phi_1}\ket{1} \}$ and (c) $\{ \alpha_0\ket{00}+\beta_0\ket{11},  \alpha_1\ket{01}+\beta_1\ket{10} \}$, where $\ket{\psi_0}, \ket{\psi_1}, \ket{\phi_0},\ket{\phi_1} $ are quantum states of single qubit and 
$\alpha_0, \alpha_1, \beta_0, \beta_1$ are complex numbers. Only form (a) and (b) are used in MQAOA.  An approximate solution for form (a) can be obtained by the following steps, and likewise for form (b).
\begin{enumerate}
	\item The density matrix $\rho^{(k)}$ is partitioned into $4$ $2\times 2$ blocks $\rho^{(k)}=\begin{bmatrix} A_{00} & A_{01} \\
		A_{10}  & A_{11} \end{bmatrix}$.
	\item $A_{00}$ and $A_{11}$ are diagonalized, $A_{00}=U_0D_0U_0^\dagger$ ,
	$A_{11}=U_1D_1U_1^\dagger$,  where $D_0, D_1$ are diagonal matrices,  $D_0=\text{diag}\{d_{00}, d_{01}\}$, $D_1=\text{diag}\{d_{10}, d_{11}\}$ with $d_{00}\le d_{01},  d_{10}\le d_{11}$.
	\item  $\rho^{(k)}$ is transformed into
	 $\tilde{\rho}^{(k)}=\begin{bmatrix} D_0 & U_0^\dagger A_{01}U_1 \\
		U_1^\dagger A_{10}U_0  & D_1 \end{bmatrix}$ by unitary operation  $\begin{bmatrix} U_0^\dagger & 0 \\
		0  & U_1^\dagger \end{bmatrix}$. The singular value decomposition of the off diagonal block
	is $U_0^\dagger A_{01}U_1=U\text{diag}\{s_0, s_1\}V$, with $s_0>s_1$. The first column of $U$ is
	denoted by $\ket{b_0}$, and the hermitian conjugate of the first row of $V$ is denoted by $\ket{b_1}$.  The first component of  both $\ket{b_0}$ and $\ket{b_1}$ are  taken as real numbers.
	\item To construct 2-dimensional subspace of $\tilde{\rho}^{(k)}$ with basis $\{ \ket{0}\ket{\tilde{\psi}_0}, \ket{1}\ket{\tilde{\psi}_1}\}$,  the following   linear combinations of
	vectors are taken as an approximate solution:    $\ket{\tilde{\psi}_0}=(d_{01}-d_{00})\ket{a_0}+(s_0-s_1)\ket{b_0}$,  $\ket{\tilde{\psi}_1}=(d_{11}-d_{10})\ket{a_0}+(s_0-s_1)\ket{b_1}$, where $\ket{a_0}=\begin{bmatrix} 0 \\1
	  \end{bmatrix}$. They are then  normalized.
	\item $\ket{\psi_0}$ and $\ket{\psi_1}$ are obtained by unitary transformations:  $\ket{\psi_0}=U_0\ket{\tilde{\psi}_0}$, $\ket{\psi_1}=U_1\ket{\tilde{\psi}_1}$.
\end{enumerate} 
In the loop of MQAOA, we alternate between using form (a) and (b), except in the first round,  where we choose the one with smaller $\left\Vert\rho^{(k)}-P\rho^{(k)} P\right\Vert$.

\end{document}